\documentclass[aps,prl,twocolumn,groupedaddress,floatfix,showpacs,showkeys,amsmath,amssymb]{revtex4-1}
\usepackage{graphicx}
\usepackage{bm}
\begin{document}
\title{Modeling Electronic Quantum Transport with Machine Learning}

\author{Alejandro Lopez-Bezanilla$^{1}$}
\email[]{alejandrolb@gmail.com}
\author{O. Anatole von Lilienfeld$^{2,3}$}
\email[]{anatole.vonlilienfeld@unibas.ch}
\affiliation{$^{1}$Materials Science Division, Argonne National Laboratory, 9700 S. Cass Avenue, Lemont, IL 60439, USA}
\affiliation{$^{2}$Argonne Leadership Computing Facility, Argonne National Laboratory, 9700 S. Cass Avenue, Lemont, IL 60439, USA}
\affiliation{$^{3}$Institute of Physical Chemistry, Department of Chemistry, University of Basel, Klingelbergstrasse 80, CH-4056 Basel, Switzerland}
\date{\today}

\begin{abstract}
We present a Machine Learning approach to solve electronic quantum transport equations of one-dimensional nanostructures. The transmission coefficients of disordered systems were computed to provide training and test datasets to the machine. The system's representation encodes energetic as well as geometrical information to characterize similarities between disordered configurations, while the Euclidean norm is used as a measure of similarity. Errors for out-of-sample predictions systematically decrease with training set size, enabling the accurate and fast prediction of new transmission coefficients. The remarkable performance of our model to capture the complexity of interference phenomena lends further support to its viability in dealing with transport problems of undulatory nature.

\end{abstract}

\maketitle
Substantial advances in computational science capabilities have opened new research frontiers, greatly expanding the impact of the material sciences community's work. The unrelenting drive towards the use of atomistic simulation for the routine generation of large data sets is obtaining great interest \cite{Curtarolo2013}. Large scale efforts such as the United States Materials Genome Initiative\cite{MaterialsProject} are aiming for the discovery and development of new compounds thanks to increasingly faster and cheaper computational resources\cite{ISI:000315707200001}. In parallel, the development of data mining techniques for intelligent interrogation of large databases are succeeding in recognizing meaningful patterns in structured data \cite{MachineLearningHautierCeder2010, Wang}. It still remains an open question, however, as to how to integrate into explicit structure-property relationships the knowledge hidden, yet implicitly present, in the data. 
Machine Learning (ML), the ability of computer algorithms to comprehend data and infer new results for new situations, is gaining importance as a tool of choice to analyze the growing and complex data generated in many scientific and engineering contexts \cite{HasTibFri01,MueMikRaeTsuSch01}. By appropriately estimating pairwise distances in a data set, supervised learning techniques directly allow for the resolution of computationally expensive sets of equations by making sense of accumulated knowledge.
Within atomistic simulation, ML already demonstrated its usefulness in predicting outcomes from known patterns and inferring new knowledge. Examples include 
chemical binding \cite{RuppPRL2012}, electronic levels \cite{Montavon2013}, and one-dimensional orbital free density functionals \cite{ML4Kieron2012}.

In this paper we use ML to develop a new and alternative {\em Ansatz} for modeling transmission coefficients of disordered one-dimensional device channels. 
This resulting technique allows us to estimate the conductivity of a large dataset of model device channels accurately, with very moderate computational effort. 
The model introduced predicts a real-valued function (electron transmission) for independent values (electron energy) based on training for examples that were previously constructed by randomly selecting a set of disordered systems. We validate the expected computed function with a second set of reference results. 
The ML model is able to capture the complex behavior of reflective electron waves canceling each other as a consequence of destructive interference upon multiple reflections between channel impurities. 
As such, this statistical model retains the underlying quantum features of the training data, and projects them into the validating set with high accuracy.

\begin{figure}[htp]
 \centering
 \includegraphics[width=0.45 \textwidth]{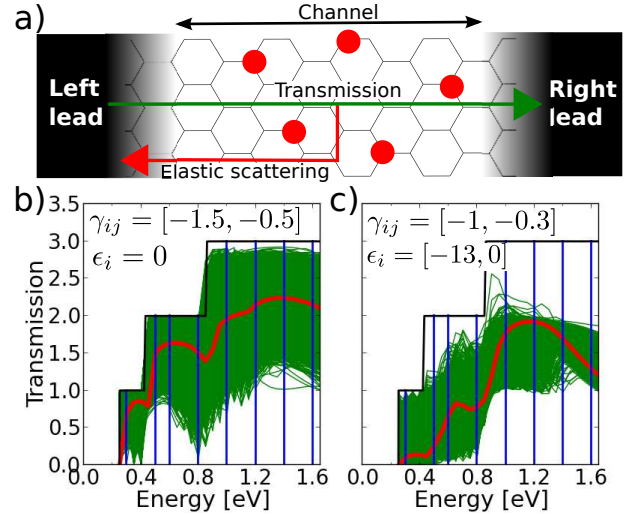}
 \caption{a) schematic diagram of the modeled system. A hexagonal network of atoms divided in three parts: left and right leads, and a central region containing scattering centers. b) and c) show the transmissions of defectless channels (black lines), and of 8000 defective configurations for 5 spatially fixed scattering centers with the parameters in inset (green lines). Red lines indicate the average transmissions, and vertical blue lines the energies at which the machine has been trained for results in Figure \ref{fig2} and Figure \ref{fig3}.}
 \label{fig1}
\end{figure}

We simulate electron transport for a conducting channel model which is assumed to be sufficiently transferable. 
The system consists of an infinite-long hexagonal network of single-orbital atoms (simple model for a graphene nanoribbon) divided in three regions (see  Figure \ref{fig1}-a). A central region (channel) exhibiting geometrical or compositional disorder is coupled to the left and right to two semi-infinite and multi-transverse mode ballistic leads. All backscattering phenomena occur in the channel which matches the leads with reflectionless contacts. 

A standard first-neighbor tight-binding Hamiltonian parametrizes energetical description of the disordered system,
\begin{equation}
H= \sum_{i} \epsilon_i |i \rangle \langle i| + \sum_{i,j}\gamma_{ij}|i\rangle \langle j|,
\end{equation}
where $\epsilon_i$ is the onsite energy of site $i$, and $\gamma_{ij}$ is the hopping element to a nearest neighbor site $j$ in the lattice.
In order to analyze the two-terminal transport through the conducting system, we use the Landauer-B\"uttiker (LB) approach \cite{INSPEC:1957A08595,INSPEC:143734,PhysRevLett.57.1761} that provides a conceptually simple framework to describe the physics of electron coherent transport at the nanoscale. In a two-probe system, the conductance is quantized in $G_0=e^2/h$, the quantum unit of conductance, and reads:
\begin{equation}
G(E)=G_0 T(E) = G_0\sum_{n=1}^NT_n(E)
\end{equation}
where the transmission coefficients $T(E)$ can be expressed as a sum over all the $N$ transmitting modes at energy $E$, and give the probability of an electron to be transmitted from one electrode to the opposite when it quantum mechanically interferes with the channel impurities. We evaluate the retarded (advanced) Green's functions of the system within the standard Green's function formalism, 
\begin{equation}
\mathcal{G}^{\pm}(E)=\{E I-H-\Sigma^{\pm}_{L}(E)-\Sigma^{\pm}_{R}(E)\}^{-1}
\end{equation}
where $\Sigma^{\pm}_{L(R)}(E)$ are the self-energies which describe the coupling of the channel to the left (-) and right (+) electrodes. These quantities are related to the transmission factor by the relation \cite{PhysRevB.23.6851}, 
\begin{equation}
T(E)=\rm{tr}\{\Gamma_{L}(E) \mathcal{G}^{+}(E) \Gamma_{R}(E) \mathcal{G}^{-}(E)\},
\label{eqtransm}
\end{equation}
with 
$\Gamma_{L(R)}(E)=i\{\Sigma^{+}_{L(R)}(E)-\Sigma^{-}_{L(R)}(E)\}$. We have implemented and applied this framework for the generation of the various test and training sets outlined below. 

An efficient supervised learning scheme relies upon a proper definition of a measure of similarity between systems. 
The probably most crucial step consists of finding a suitable representation, also known as ``descriptor'' \cite{FourierDescriptor}, 
which should fulfill certain requirements such as symmetry invariance, uniqueness, or differentiability. 
Our electron transport model relies on the following square matrix $\bf{M}$ as a descriptor,
\begin{eqnarray}
M_{ij}  =
\begin{cases}
 \epsilon_i & \forall \;\; i = j,\\
  \gamma_{ij} & \forall \;\; i \;\; {\rm adjacent} \;\; j,\\
   \sqrt{d_{ij}} & {\rm elsewise}
\end{cases}
\label{eq:matrix}
\end{eqnarray}
where $d_{ij}=|{\bf R}_i - {\bf R}_j|$ is the distance between two sites. 
The dimensionality of $\bf{M}$ is the number of atoms in the channel. 
Invariance with respect to site indexing is enforced by sorting the atom site indices according to the norm of the rows of $\bf{M}$.  
Note that $\bf{M}$ encodes the system's identity in terms of energetic descriptions and geometrical configurations, the same information that also defines the Hamiltonian entering the Green's function formalism.

For the ML model we rely on the standard Laplacian kernel model that has already been used in the context of a wide range of applications \cite{AssessmentMLJCTC2013}. 
This ML model estimates the transmission $T$ at energy $E$ for a system with descriptor ${\bf M}_J$ as a sum of weighted exponential functions, 
\begin{equation}
T^{\rm est}(E,{\bf M}_J)= \sum_{I=1}^{N_t} \alpha_I(E) e^{-\frac{D_{IJ}}{\sigma}},
\label{eq:T}
\end{equation}
where $N_t$ is the number of samples in the training set, and where $D_{IJ} = |{\bf M}_I - {\bf M}_J|$, i.e.~the Euclidean norm between two channels $I$ and $J$.
The regression coefficients $\{\alpha_I(E)\}$, and length-scale $\sigma(E)$, are obtained at discrete values of $E$ through kernel ridge regression, 
${\bm \alpha}(E) = [{\bf K}-\lambda {\bf I}]^{-1}{\bf T}^{\rm ref}(E)$,
where $\bf{K}$ is the kernel matrix with elements $K_{IJ}=e^{-D_{IJ}/\sigma}$, and where ${\bf T}^{\rm ref}(E)$ is the vector of reference transmission coefficients at $E$.  
Since the employed reference data is noise-free, the regularization parameter $\lambda$, ordinarily used to account for the noise in experimental data~\cite{AssessmentMLJCTC2013}, has been set to zero.
% 0
The characteristic length that yield the best model's performance for all runs was $\sigma$ = 1000. 
\begin{figure*}[htp]
 \centering
 \includegraphics[width=0.9 \textwidth]{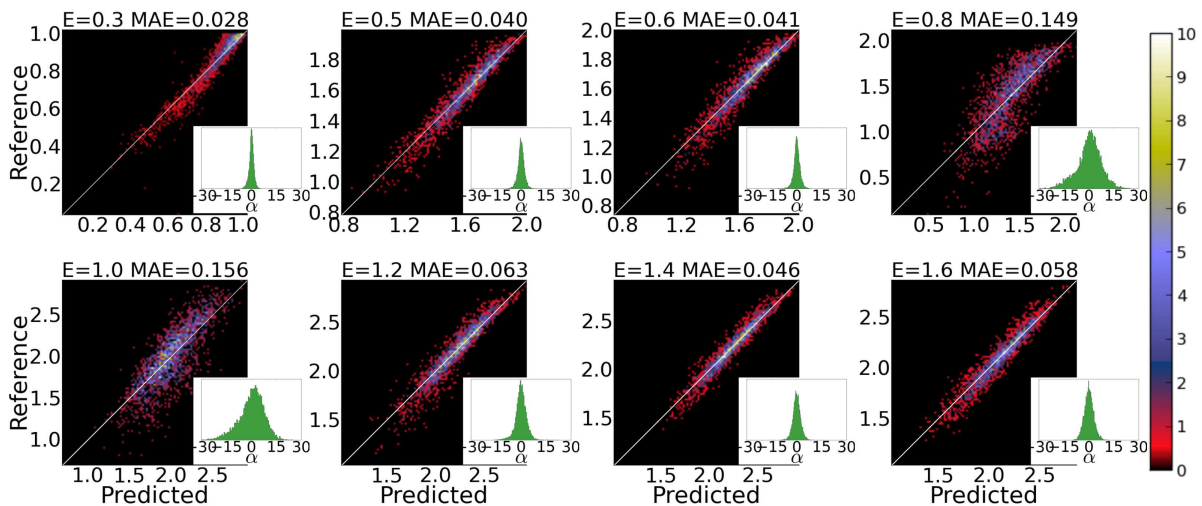}
 \caption{ 2D histograms of correlated 2000 reference transmissions with machine learning-based predicted values, for $N_t = $ 8000 samples with 5 spatially fixed scattering centers and parameters $\epsilon_i=0$ and $\gamma_{ij}\in[-1.5,-0.5]$. Above each panel the energy E at which the transmissions were computed, and the mean absolute error (MAE), are indicated. Insets show the histograms of the corresponding $\{\alpha_I(E)\}$ distributions.}
 \label{fig2}
\end{figure*}
Transport calculations adopting the aforementioned Green's function formulation have been used to generate several training and testing data sets. 
The impurities are randomly scattered along a channel formed by repeating 4 times an armchair graphene nanoribbon unit cell with 7 single-orbital atom dimers across the ribbon width.

First, we discuss ML results for a training set consisting of $N_t = $ 8000 different samples, all with the same five impurity sites chosen at random, as also illustrated in Figure (\ref{fig1}-a).
Hopping terms of the impurities to their nearest neighbors have been set randomly within the range $\gamma_{ij}\in[-1.5,-0.5]$. 
For the rest of sites in the leads and the channel, $\gamma_{ij}=-1$. 
Onsite energies have been set to $\epsilon_i=0$ for all sites. 
The black line in Figure \ref{fig1}-b illustrates the stepwise increasing transmission of a defectless channel for increasing values of $E$ in the conduction band. Green lines show for defective channels the transmission drop as result of the backscattering, and represent the extent of the transmission variation for the considered range of energy. 
For various discrete values of $E$, highlighted with vertical blue lines in Figure \ref{fig1}-b, 
we have trained a ML model and tested its performance for 2000 ``out-of-sample'' defected channels. 
Figure \ref{fig2} shows the predictive performance in various panels, each corresponding to a different energy value, in heat histogram representation.
Overall, the correlations between the ML (predicted) and Equation \ref{eqtransm} (reference) results are remarkable, 
with errors routinely scoring at less than 5\% of the average transmission coefficients.
Notice that the accuracy of the model varies as a function of electron energy, being the largest error at $E$ = 0.8 and 1 eV. 
This is also manifested by the corresponding $\{\alpha_I(E)\}$ histograms (insets of Figure \ref{fig2}), which evince a direct relation between the broadening of the computed coefficients of Equation \ref{eq:T} and the quality of the prediction. It is observed that $\alpha$ varies smoothly as a function of $E$, pointing out that the contribution of a given sample does not change abruptly as $E$ is tuned. This might help to improve the model's accuracy in the future through inclusion of derivatives of $\alpha$ with respect to $E$.

The two peaks in MAE are clearly consistent with the apparition of new transmission modes which occur in the vicinity of $E$ = 0.4 and 0.8 eV. 
The wide variability of the transmission coefficients from sample to sample at a resonance energy and in the vicinity of a van Hoff singularity hinder the training process and, thereby, enhance the unpredictability. This can be ascribed to the interference phenomena resulting from the multiple reflections of electron waves with the scattering centers, rendering $T(E)$ strongly sensitive to both the distance between impurities and the strength of $\gamma_{ij}$. Notice the linear decrease of the MAE with logarithm of training set size for each $E$ value, including those of more difficult predictability. 
\begin{figure}[htp]
 \centering
  \includegraphics[width=0.45 \textwidth]{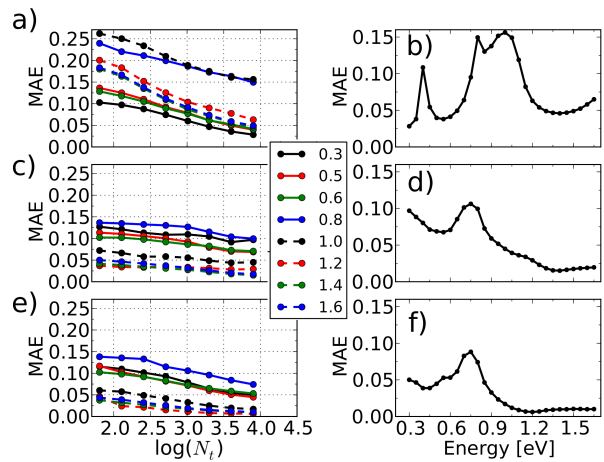}
 \caption{Linear drop of the mean absolute error (MAE) as a function of the logarithm of the number of training samples ($N_t$), in a) for $\epsilon_i=0$ and $\gamma_{ij}\in[-1.5,-0.5]$, in c) for $\epsilon_i\in[-13,0]$ and $\gamma_{ij}\in[-1,-0.3]$, and in e) with the same parameters as in the latter after removal of 10\% of the outliers. b), d) and f) show the MAE as a function of the energy for $N_t = $ 8000 samples.}
 \label{fig3}
\end{figure}

In a second experiment, we generated more complex training and test sets introducing an additional source of disorder. The scattering efficiency is enhanced by allowing the $\epsilon_i$ of the impurities to take finite values. 
Figure \ref{fig1}-c) features the corresponding transmission profiles for $\epsilon_i\in$[-13,0] and $\gamma_{ij}\in$[-1,-0.3] for another training set with $N_t = $ 8000 samples. 
Fig. \ref{fig3}-b) illustrates again a linear drop of the MAE with training set size. 
Despite the additional disorder, the higher degree of localization induced by the activation of the onsite energies leads to a significantly reduced MAE.
For larger $N_t$, however, both models converge to similar error ranges.
The $E$ dependency of the MAE is moderated in this case, exhibiting only a small peak at $E$ = 0.8 eV, in the close vicinity to a new transmitting mode onset.
Because small sample to sample variations may yield large differences in the transmission, some training data may be difficult to classify. This usually involves additional effort for the learning task, either in debasing the performance, or in slowing down the error decreasing with the training set size. A reduction of the variation range through the removal of 10\% of outliers whose $T(E)$ deviate most from the average supports this surmise. Figures \ref{fig3}-e and f show a significant improvement of the predictive power with reductions in the MAE by up to 50 \%. 

\begin{figure}[htp]
 \centering
 \includegraphics[width=0.5 \textwidth]{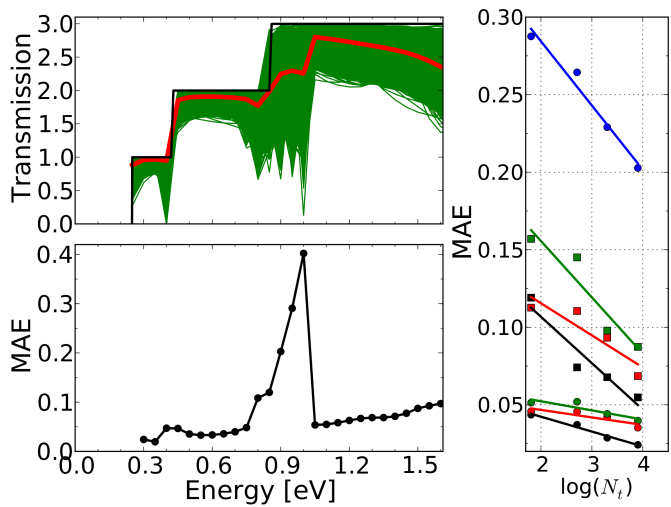}
 \caption{a) Transmissions of $N_t = $ 8000 defective channels (green lines) with various $\gamma_{ij}$ strength and changeable random positions. Red line indicates the average.
%  Black line indicates the transmission of a pristine channel. 
b) shows the MAE as a function of the energy for the $N_t$ samples. c) Linear drop of the mean absolute error (MAE) as a function of the logarithm of $N_t$ in a). }
 \label{fig4}
\end{figure}

Thirdly, we investigate the ML model resulting from a training set with variable random positions of the scattering centers in combination with random $\gamma_{ij}$ strength. 
Specifically, we generated 100 samples with different random positions. 
From each of these samples we generated 100 defected configurations with $\gamma_{ij}$ $\in$[-1,-0.7] randomly chosen for the impurity sites, and $\gamma_{ij}$ = -1 eV for all other sites.  $\epsilon_i$ has been set to 0 throughout. 8000 out of the resulting 10000 channels have been selected at random to train our ML model, and the remaining 2000 were used for out-of-sample testing. 
Green lines of Figure \ref{fig4}-a features the resulting transmission coefficients for this training set. The resulting evolution plot of the MAE with $N_t$ in Figure \ref{fig4}-b demonstrates a very good correlation with the predicted results. Only in the proximity of the resonance at $E$ = 1 eV, however, the variation in $T$ becomes large, the MAE increasing by an order of magnitude to $\sim$ 0.4. Despite such a large error, Figure \ref{fig4}-c clearly suggests that even at the resonance energy a lower error can be achieved through extension of the training set size. 

To conclude, we have introduced a ML model for predicting electronic quantum transmission coefficients as a function of electron energy for one-dimensional channels. 
Our numerical results suggest that the model is capable of integrating previously computed transmission coefficient data into a 
simple and efficient framework, and of inferring transmission coefficients for new (out-of-sample) channels. 
The proposed descriptor has proved to be highly efficient in encoding the defected channels' identity. 
The remarkable performance of this ML scheme when it comes to capture the complexity of interference phenomena 
lends further support to its viability in dealing with transport problems of undulatory nature.
Furthermore, as follows from the different complexity of the equations to solve, the ML model is dramatically less computationally demanding than conventional models and, given a sufficiently large training set of disordered channels, yields competitive accuracy %(see Supplemental Materials for details). 
% (In the Supplementary Information a benchmark analysis of GF vs ML approach performance is presented in detail).
In summary, we have shown that non-linear statistical regression approaches offer promising alternatives for solving the electron transmission problem in disordered nanostructures.

This research used resources of the Argonne Leadership Computing Facility at Argonne National Laboratory,
which is supported by the Office of Science of the U.S.~DOE under contract DE-AC02-06CH11357.
OAvL acknowledges funding from the Swiss National Science foundation (No.~PPOOP2\_ 138932).

% \bibliography{literatur}{1}  

\begin{thebibliography}{15}
\expandafter\ifx\csname natexlab\endcsname\relax\def\natexlab#1{#1}\fi
\expandafter\ifx\csname bibnamefont\endcsname\relax
  \def\bibnamefont#1{#1}\fi
\expandafter\ifx\csname bibfnamefont\endcsname\relax
  \def\bibfnamefont#1{#1}\fi
\expandafter\ifx\csname citenamefont\endcsname\relax
  \def\citenamefont#1{#1}\fi
\expandafter\ifx\csname url\endcsname\relax
  \def\url#1{\texttt{#1}}\fi
\expandafter\ifx\csname urlprefix\endcsname\relax\def\urlprefix{URL }\fi
\providecommand{\bibinfo}[2]{#2}
\providecommand{\eprint}[2][]{\url{#2}}

\bibitem[{\citenamefont{Curtarolo et~al.}(2013)\citenamefont{Curtarolo, Hart,
  Nardelli, Mingo, Sanvito, and Levy}}]{Curtarolo2013}
\bibinfo{author}{\bibfnamefont{S.}~\bibnamefont{Curtarolo}},
  \bibinfo{author}{\bibfnamefont{G.~L.~W.} \bibnamefont{Hart}},
  \bibinfo{author}{\bibfnamefont{M.~B.} \bibnamefont{Nardelli}},
  \bibinfo{author}{\bibfnamefont{N.}~\bibnamefont{Mingo}},
  \bibinfo{author}{\bibfnamefont{S.}~\bibnamefont{Sanvito}}, \bibnamefont{and}
  \bibinfo{author}{\bibfnamefont{O.}~\bibnamefont{Levy}},
  \bibinfo{journal}{Nature Mater} \textbf{\bibinfo{volume}{12}},
  \bibinfo{pages}{191} (\bibinfo{year}{2013}).

\bibitem[{\citenamefont{Ong et~al.}(2011)\citenamefont{Ong, Jain, Hautier,
  Kocher, Cholia, Gunter, Bailey, Skinner, Persson, and
  Ceder}}]{MaterialsProject}
\bibinfo{author}{\bibfnamefont{S.~P.} \bibnamefont{Ong}},
  \bibinfo{author}{\bibfnamefont{A.}~\bibnamefont{Jain}},
  \bibinfo{author}{\bibfnamefont{G.}~\bibnamefont{Hautier}},
  \bibinfo{author}{\bibfnamefont{M.}~\bibnamefont{Kocher}},
  \bibinfo{author}{\bibfnamefont{S.}~\bibnamefont{Cholia}},
  \bibinfo{author}{\bibfnamefont{D.}~\bibnamefont{Gunter}},
  \bibinfo{author}{\bibfnamefont{D.}~\bibnamefont{Bailey}},
  \bibinfo{author}{\bibfnamefont{D.}~\bibnamefont{Skinner}},
  \bibinfo{author}{\bibfnamefont{K.~A.} \bibnamefont{Persson}},
  \bibnamefont{and} \bibinfo{author}{\bibfnamefont{G.}~\bibnamefont{Ceder}},
  \emph{\bibinfo{title}{{The Materials Project}}} (\bibinfo{year}{2011}).

\bibitem[{\citenamefont{Curtarolo}({2013})}]{ISI:000315707200001}
\bibinfo{author}{\bibfnamefont{S.}~\bibnamefont{Curtarolo}},
  \bibinfo{journal}{{naturematerials}} \textbf{\bibinfo{volume}{{12}}},
  \bibinfo{pages}{{173}} (\bibinfo{year}{{2013}}).

\bibitem[{\citenamefont{Hautier et~al.}(2010)\citenamefont{Hautier, Fischer,
  Jain, Mueller, and Ceder}}]{MachineLearningHautierCeder2010}
\bibinfo{author}{\bibfnamefont{G.}~\bibnamefont{Hautier}},
  \bibinfo{author}{\bibfnamefont{C.~C.} \bibnamefont{Fischer}},
  \bibinfo{author}{\bibfnamefont{A.}~\bibnamefont{Jain}},
  \bibinfo{author}{\bibfnamefont{T.}~\bibnamefont{Mueller}}, \bibnamefont{and}
  \bibinfo{author}{\bibfnamefont{G.}~\bibnamefont{Ceder}},
  \bibinfo{journal}{Chem. Mater.} \textbf{\bibinfo{volume}{22}},
  \bibinfo{pages}{3762} (\bibinfo{year}{2010}).

\bibitem[{\citenamefont{Pilania et~al.}(2013)\citenamefont{Pilania, Wang,
  Jiang, Rajasekaran, and Ramprasad}}]{Wang}
\bibinfo{author}{\bibfnamefont{G.}~\bibnamefont{Pilania}},
  \bibinfo{author}{\bibfnamefont{C.}~\bibnamefont{Wang}},
  \bibinfo{author}{\bibfnamefont{X.}~\bibnamefont{Jiang}},
  \bibinfo{author}{\bibfnamefont{S.}~\bibnamefont{Rajasekaran}},
  \bibnamefont{and}
  \bibinfo{author}{\bibfnamefont{R.}~\bibnamefont{Ramprasad}},
  \bibinfo{journal}{Sci. Rep.} \textbf{\bibinfo{volume}{3}}
  (\bibinfo{year}{2013}).

\bibitem[{\citenamefont{Hastie et~al.}(2001)\citenamefont{Hastie, Tibshirani,
  and Friedman}}]{HasTibFri01}
\bibinfo{author}{\bibfnamefont{T.}~\bibnamefont{Hastie}},
  \bibinfo{author}{\bibfnamefont{R.}~\bibnamefont{Tibshirani}},
  \bibnamefont{and} \bibinfo{author}{\bibfnamefont{J.}~\bibnamefont{Friedman}},
  \emph{\bibinfo{title}{The Elements of Statistical Learning: data mining,
  inference and prediction}}, Springer series in statistics
  (\bibinfo{publisher}{Springer}, \bibinfo{address}{New York, N.Y.},
  \bibinfo{year}{2001}).

\bibitem[{\citenamefont{M\"uller et~al.}(2001)\citenamefont{M\"uller, Mika,
  R{\"a}tsch, Tsuda, and Sch{\"o}lkopf}}]{MueMikRaeTsuSch01}
\bibinfo{author}{\bibfnamefont{K.-R.} \bibnamefont{M\"uller}},
  \bibinfo{author}{\bibfnamefont{S.}~\bibnamefont{Mika}},
  \bibinfo{author}{\bibfnamefont{G.}~\bibnamefont{R{\"a}tsch}},
  \bibinfo{author}{\bibfnamefont{K.}~\bibnamefont{Tsuda}}, \bibnamefont{and}
  \bibinfo{author}{\bibfnamefont{B.}~\bibnamefont{Sch{\"o}lkopf}},
  \bibinfo{journal}{IEEE Transactions on Neural Networks}
  \textbf{\bibinfo{volume}{12}}, \bibinfo{pages}{181} (\bibinfo{year}{2001}).

\bibitem[{\citenamefont{Rupp et~al.}(2012)\citenamefont{Rupp, Tkatchenko,
  M\"uller, and von Lilienfeld}}]{RuppPRL2012}
\bibinfo{author}{\bibfnamefont{M.}~\bibnamefont{Rupp}},
  \bibinfo{author}{\bibfnamefont{A.}~\bibnamefont{Tkatchenko}},
  \bibinfo{author}{\bibfnamefont{K.-R.} \bibnamefont{M\"uller}},
  \bibnamefont{and} \bibinfo{author}{\bibfnamefont{O.~A.} \bibnamefont{von
  Lilienfeld}}, \bibinfo{journal}{Phys. Rev. Lett.}
  \textbf{\bibinfo{volume}{108}}, \bibinfo{pages}{058301}
  (\bibinfo{year}{2012}).

\bibitem[{\citenamefont{Montavon et~al.}(2013)\citenamefont{Montavon, Rupp,
  Gobre, Vazquez-Mayagoitia, Hansen, Tkatchenko, Müller, and von
  Lilienfeld}}]{Montavon2013}
\bibinfo{author}{\bibfnamefont{G.}~\bibnamefont{Montavon}},
  \bibinfo{author}{\bibfnamefont{M.}~\bibnamefont{Rupp}},
  \bibinfo{author}{\bibfnamefont{V.}~\bibnamefont{Gobre}},
  \bibinfo{author}{\bibfnamefont{A.}~\bibnamefont{Vazquez-Mayagoitia}},
  \bibinfo{author}{\bibfnamefont{K.}~\bibnamefont{Hansen}},
  \bibinfo{author}{\bibfnamefont{A.}~\bibnamefont{Tkatchenko}},
  \bibinfo{author}{\bibfnamefont{K.-R.} \bibnamefont{Müller}},
  \bibnamefont{and} \bibinfo{author}{\bibfnamefont{O.~A.} \bibnamefont{von
  Lilienfeld}}, \bibinfo{journal}{New Journal of Physics}
  \textbf{\bibinfo{volume}{15}}, \bibinfo{pages}{095003}
  (\bibinfo{year}{2013}).

\bibitem[{\citenamefont{Snyder et~al.}(2012)\citenamefont{Snyder, Rupp, Hansen,
  M\"uller, and Burke}}]{ML4Kieron2012}
\bibinfo{author}{\bibfnamefont{J.~C.} \bibnamefont{Snyder}},
  \bibinfo{author}{\bibfnamefont{M.}~\bibnamefont{Rupp}},
  \bibinfo{author}{\bibfnamefont{K.}~\bibnamefont{Hansen}},
  \bibinfo{author}{\bibfnamefont{K.-R.} \bibnamefont{M\"uller}},
  \bibnamefont{and} \bibinfo{author}{\bibfnamefont{K.}~\bibnamefont{Burke}},
  \bibinfo{journal}{Phys. Rev. Lett.} \textbf{\bibinfo{volume}{108}},
  \bibinfo{pages}{253002} (\bibinfo{year}{2012}).

\bibitem[{\citenamefont{Landauer}(1957)}]{INSPEC:1957A08595}
\bibinfo{author}{\bibfnamefont{R.}~\bibnamefont{Landauer}},
  \bibinfo{journal}{IBM Journal of Research and Development}
  \textbf{\bibinfo{volume}{1}}, \bibinfo{pages}{223} (\bibinfo{year}{1957}).

\bibitem[{\citenamefont{Landauer}(1970)}]{INSPEC:143734}
\bibinfo{author}{\bibfnamefont{R.}~\bibnamefont{Landauer}},
  \bibinfo{journal}{Philosophical Magazine} \textbf{\bibinfo{volume}{21}},
  \bibinfo{pages}{863} (\bibinfo{year}{1970}).

\bibitem[{\citenamefont{B\"uttiker}(1986)}]{PhysRevLett.57.1761}
\bibinfo{author}{\bibfnamefont{M.}~\bibnamefont{B\"uttiker}},
  \bibinfo{journal}{Phys. Rev. Lett.} \textbf{\bibinfo{volume}{57}},
  \bibinfo{pages}{1761} (\bibinfo{year}{1986}).

\bibitem[{\citenamefont{Fisher and Lee}(1981)}]{PhysRevB.23.6851}
\bibinfo{author}{\bibfnamefont{D.~S.} \bibnamefont{Fisher}} \bibnamefont{and}
  \bibinfo{author}{\bibfnamefont{P.~A.} \bibnamefont{Lee}},
  \bibinfo{journal}{Phys. Rev. B} \textbf{\bibinfo{volume}{23}},
  \bibinfo{pages}{6851} (\bibinfo{year}{1981}).

\bibitem[{\citenamefont{von Lilienfeld et~al.}(2013)\citenamefont{von
  Lilienfeld, Rupp, and Knoll}}]{FourierDescriptor}
\bibinfo{author}{\bibfnamefont{O.~A.} \bibnamefont{von Lilienfeld}},
  \bibinfo{author}{\bibfnamefont{M.}~\bibnamefont{Rupp}}, \bibnamefont{and}
  \bibinfo{author}{\bibfnamefont{A.}~\bibnamefont{Knoll}}
  (\bibinfo{year}{2013}), \bibinfo{note}{submitted to J Chem Phys}.

\bibitem [{\citenamefont {Hansen}\ \emph {et~al.}(2013)\citenamefont {Hansen},
  \citenamefont {Montavon}, \citenamefont {Biegler}, \citenamefont {Fazli},
  \citenamefont {Rupp}, \citenamefont {Scheffler}, \citenamefont {von
  Lilienfeld}, \citenamefont {Tkatchenko},\ and\ \citenamefont
  {M\"uller}}]{AssessmentMLJCTC2013}%
  \BibitemOpen
  \bibfield  {author} {\bibinfo {author} {\bibfnamefont {K.}~\bibnamefont
  {Hansen}}, \bibinfo {author} {\bibfnamefont {G.}~\bibnamefont {Montavon}},
  \bibinfo {author} {\bibfnamefont {F.}~\bibnamefont {Biegler}}, \bibinfo
  {author} {\bibfnamefont {S.}~\bibnamefont {Fazli}}, \bibinfo {author}
  {\bibfnamefont {M.}~\bibnamefont {Rupp}}, \bibinfo {author} {\bibfnamefont
  {M.}~\bibnamefont {Scheffler}}, \bibinfo {author} {\bibfnamefont {O.~A.}\
  \bibnamefont {von Lilienfeld}}, \bibinfo {author} {\bibfnamefont
  {A.}~\bibnamefont {Tkatchenko}}, \ and\ \bibinfo {author} {\bibfnamefont
  {K.-R.}\ \bibnamefont {M\"uller}},\ }\href {\doibase 10.1021/ct400195d}
  {\bibfield  {journal} {\bibinfo  {journal} {Journal of Chemical Theory and
  Computation}\ }\textbf {\bibinfo {volume} {9}},\ \bibinfo {pages} {3404}
  (\bibinfo {year} {2013})}.
\end{thebibliography}

\end{document}